\documentclass[twocolumn,prl,superscriptaddress,preprintnumbers,amsmath,amssymb]{revtex4}
\usepackage{graphicx}
\usepackage{dcolumn}
\usepackage{bm}

\newcommand{\ave}[1]{\langle {#1} \rangle}

\begin{document}
\title{Universal relaxation dynamics of sphere packings below jamming}

\author{Atsushi Ikeda}

\affiliation{Graduate School of Arts and Sciences, The University of Tokyo, Tokyo 153-8902, Japan}
\affiliation{Research Center for Complex Systems Biology, Universal Biology Institute, University of Tokyo, Komaba, Tokyo 153-8902, Japan}

\author{Takeshi Kawasaki}

\affiliation{Department of Physics, Nagoya University, Nagoya 464-8602, Japan}

\author{Ludovic Berthier}

\affiliation{Laboratoire Charles Coulomb (L2C), University of Montpellier, CNRS, Montpellier, France}

\author{Kuniyasu Saitoh}

\affiliation{Research Alliance Center for Mathematical Sciences \& WPI-Advanced Institute for Materials Research, Tohoku University, Sendai 980-8577, Japan}

\author{Takahiro Hatano}
\affiliation{Department of Earth and Space Science, Osaka University, 560-0043 Osaka, Japan}

\date{\today}
\begin{abstract}
We show that non-Brownian suspensions of repulsive spheres below jamming display a slow relaxational dynamics with a characteristic time scale that diverges at jamming. This slow time scale is fully encoded in the structure of the unjammed packing and can be readily measured via the vibrational density of states. We show that the corresponding dynamic critical exponent is the same for randomly generated and sheared packings. Our results show that a wide variety of physical situations, from suspension rheology to algorithmic studies of the jamming transition are controlled by a unique diverging timescale, with a universal critical exponent. 
\end{abstract}

\maketitle

{\em Introduction.}--- 
Disordered particle packings solidify when they are compressed~\cite{Liu1998,Larson1999,Coussot2005,Berthier2011}. 
In the absence of thermal motion~\cite{Ikeda2012,Olsson2013,Ikeda2013,Dinkgreve2018}, as in non-Brownian suspensions~\cite{Boyer2011}, emulsions of large droplets~\cite{Mason1995}, foams~\cite{Bolton1990}, and granular materials~\cite{Jaeger1996}, this phenomenon is a jamming transition~\cite{Liu1998}. Athermal frictionless spheres~\cite{Durian1995,Ohern2003} represent an idealized model to study the jamming criticality~\cite{DH,vanHecke2011}. At the critical packing fraction $\varphi = \varphi_J$, the system is isostatic~\cite{Ohern2003}, the contact number is $z = 2d$ where $d$ is the spatial dimension, and the system is mechanically marginally stable~\cite{Alexander1998}. In the jammed phase $\varphi > \varphi_J$, the mechanical response~\cite{Ohern2003,Ellenbroek2006,Tighe2011,Lerner2014,Karimi2015,Mizuno2016} and low-frequency vibrational modes~\cite{Silbert2005,Ikeda2013a,Charbonneau2016,Mizuno2017} exhibit algebraic critical behaviour. Theoretical developments~\cite{Wyart2005,Yan2016}, effective medium theory~\cite{Wyart2010,DeGiuli2014}, and an exact analytic solution in the $d \to \infty$ mean-field limit~\cite{Parisi2010,Charbonneau2014,Franz2015} provide a good understanding of the jammed phase, including numerical values of critical exponents. 

By contrast, current understanding of the unjammed phase $\varphi < \varphi_J$ is more limited, in particular regarding dynamics.
An important physical observable in this regime is the diverging shear viscosity of non-Brownian suspensions, $\eta \propto (\varphi_J - \varphi)^{-\nu}$~\cite{Olsson2007,Boyer2011,Forterre2008,Hatano2008}. 
Even for idealised frictionless spheres~\cite{Olsson2007,Lerner2012}, an exact mean-field prediction for $\nu$ is not available, but several distinct predictions exist: $\nu \approx 2.83$ in \cite{Lerner2012a,During2014,DeGiuli2015} and $\nu =2$ in \cite{Suzuki2015}. Measured numerical values are spread in the range $\nu \in [2,4]$~\cite{Olsson2007,Nordstrom2010,Boyer2011,Lerner2012,Andreotti2012,Kawasaki2015,Suzuki2015,Olsson2015,Olsson2018}. 
We argue that physical situations where dynamics of unjammed packings is relevant go well beyond steady shear rheology. In emulsions, nearly jammed packings are prepared by dispersing the droplets in a solvent and performing centrifugation or creaming under gravity~\cite{Mason1995,Brujic2003,Jorijadze2013}, during which droplets relax. Foams near jamming are also prepared by foaming solutions and injecting the bubbles into the sample chamber~\cite{Hertzhft2005,Katgert2008,Kurita2017}, during which bubbles can relax. These materials can also be probed rheologically by performing step stress or strain experiments, monitoring the mechanical response. In simulations, packings are often prepared by relaxing a random configuration to mechanical equilibrium~\cite{Ohern2003}. In the unjammed phase $\varphi < \varphi_J$, relaxation stops when particles are just touching, and $z < 2d$~\cite{Morse2014}. Critical slowing down of relaxation algorithms near jamming was reported~\cite{torquato}. 

Several pieces of information are known from the analysis of computer models. The diverging shear viscosity is not associated to a slowing down of particle diffusion, as would be the case in equilibrium due to Green-Kubo relations~\cite{Hansen1986}. In fact, particle motion accelerates near jamming~\cite{Olsson2010,Heussinger2010}. Instead, it was shown 
that the relaxation dynamics after sudden shear cessation is characterised by a time scale $\tau$ that diverges as $\varphi_J$ is approached~\cite{Olsson2015}, with the observation that $\eta \propto \tau \propto (\Delta z)^{-3.7}$ in $d=3$, where $\Delta z = 2d -z$~\cite{Olsson2018}. 
Strikingly, $\nu$ seems to depend on dimensionality~\cite{Olsson2018}, unlike other critical exponent of jamming~\cite{Ohern2003} and theoretical predictions~\cite{Lerner2012a,During2014,DeGiuli2015,Suzuki2015}.
In Ref.~\cite{Lerner2012}, a dynamical matrix was constructed for unjammed hard spheres packings under shear. Its spectrum displays, in addition to more traditional signatures of soft modes near jamming, an isolated low-frequency mode at some $\omega_{\rm min} \approx \Delta z^{1.5}$ for $d=3$. This mode has no analog above jamming, and it is directly connected with the shear viscosity, $\eta \approx \omega_{\rm min}^{-2}$~\cite{Lerner2012}. A theoretical argument to explain this mode was also proposed, which predicts $\omega_{\rm min} \approx \Delta z^{1.7}$~\cite{Lerner2012a,During2014,DeGiuli2015}. Finally, Refs.~\cite{Durian1995,Hatano2009} studied the relaxation dynamics of random packings of soft particles below jamming after applying a step shear strain. The ensuing relaxation dynamics slows down algebraically as $\varphi \to \varphi_J$, with $\tau \propto (\varphi_J - \varphi)^{-3.3}$ for $d=3$~\cite{Hatano2009}. 

Here, we suggest that all these forms of relaxation dynamics below jamming exhibit a critical slowing down approaching jamming characterised by a unique diverging time scale $\tau$. Independently of the packing preparation, $\tau$ is fully encoded in the packing structure and can be readily accessed via the standard density of states, so that $\tau = 1/(2 \omega_{\min}^{2}) \propto \eta$. This diverging time scale is characterised by a universal exponent which depends neither on the form of the repulsion potential nor on the preparation protocol, $\omega_{\rm min} \propto \Delta z^{\alpha}$, with $\alpha \approx 1.6$ in $d=3$. 

{\em Model and methods.}--- 
We consider a three dimensional 50:50 binary mixture of large and small soft spheres with diameter ratio 1.4. The potential energy of the system is given by $E = \sum_{i >j} \frac{\epsilon}{2} (1-r_{ij}/\sigma_{ij})^2 \Theta(\sigma_{ij} - r_{ij})$, where $\Theta (x)$ is the Heaviside step function, $r_{ij}$ is the distance between particles $i$ and $j$, $\sigma_{ij} = (\sigma_i + \sigma_j)/2$ and $\sigma_i$ is the particle diameter of particle $i$. 
We study the athermal overdamped dynamics of the model,  
\begin{eqnarray}
\xi \frac{d \vec{r}_i}{d t} = - \frac{\partial E}{\partial \vec{r}_i}, \label{eom}
\end{eqnarray}
where $\vec{r}_i$ is the position of particle $i$, and $\xi$ is the viscous damping.
We focus on the unjammed phase $\varphi < \varphi_J$, where the system relaxes at long times into a zero-energy configuration. We use the small particle diameter $\sigma$, $\epsilon$ and $\xi \sigma^2/\epsilon$ as units of length, energy and time. 

We consider two types of initial configurations. First, we start from random initial configurations. We place $N$ spheres randomly in the box at packing fraction $\varphi$, and solve Eq.~(\ref{eom}) until $E/N \le 10^{-18}$. We prepare 1000 independent initial configurations for each packing fraction. 
Second, we use steady-shear initial configurations. We perform overdamped simple shear simulations of the same model at finite shear rate $\dot{\gamma}$ to reach steady state. For these simulations, we use the algorithm developed in Ref.~\cite{Kawasaki2015}, where the pressure $p$ and the shear rate $\dot{\gamma}$ are fixed and $\varphi$ fluctuates during the simulation. 
We sample configurations from the simulations at $p = 10^{-5}$ and $\dot{\gamma} \ge 10^{-8}$, where the shear viscosity is Newtonian. After steady state is reached, we suddenly stop the shear and again solve Eq.~(\ref{eom}) until $E/N \le 10^{-18}$. For each shear rate, which is specified by the average density $\ave{\varphi}$ in this paper, we obtain 50 independent initial configurations. We use standard periodic boundary conditions for the random case, and Lees-Edwards boundary conditions for the sheared case~\cite{Allen1987}. The number of particles is $N=3000$, unless otherwise noted. 

\begin{figure}[t]
\includegraphics[width=\columnwidth]{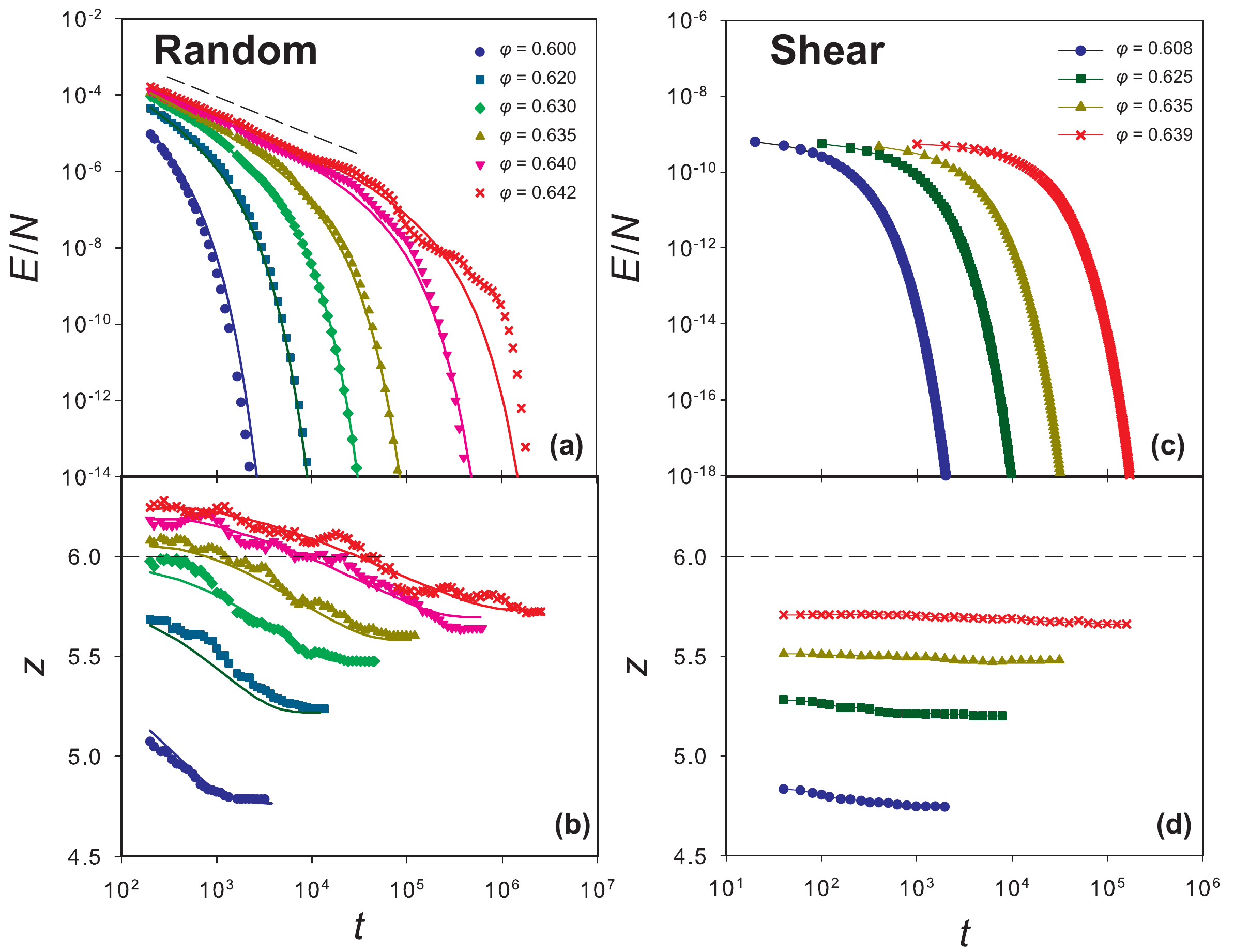}
\caption{(a,b) Time evolution of the potential energy density $E/N$ and contact number $z$ starting from random initial configurations for various packing fraction $\varphi$. Symbols and lines are typical results for a single configuration with $N=3000$ and $10^6$, respectively. 
(c,d) The same as (a,b) with initial configuration drawn from the sheared steady state with $N=3000$. Dashed line indicates $E(t) \sim t^{-1}$ in (a) and $z = 6$ in (b,d).}
\label{fig1}
\end{figure}

{\em Relaxation dynamics.}--- 
We first study the relaxation dynamics from random initial configurations. 
Figure~\ref{fig1}(a) shows typical time evolutions of the potential energy $E(t)$ at various packing fractions. It displays a simple exponential relaxation at low density, but it exhibits two time regimes at higher density with a power-law decay $t^{-1}$ at short times, followed by an exponential decay $e^{-t/\tau}$ at long times. The final relaxation time $\tau$ increases rapidly as $\varphi_J$ is approached. We simultaneously measure the time evolution of the contact number $z(t)$ in Fig.~\ref{fig1}(b)~\footnote{Time evolution of the contact numbers in Figs.~\ref{fig1}(b,d) are calculated without removing rattlers. The contact numbers shown in Figs.~\ref{fig4}, \ref{fig5}, where we performed quantitative analysis, are obtained after removing rattlers.}. It decreases with time for $t < \tau$, which implies that contact breaking and restructuring occurs on this time scale. On the other hand, $z(t)$ becomes nearly constant for $t > \tau$, which suggests that the final exponential decay of the energy takes place with a fixed contact network. 
Note that $z(t)$ approaches a finite value as $t \to \infty$, and particles are still touching in the final configurations, even though $\varphi<\varphi_J$. This is natural because in the overdamped dynamics, the particles' overlaps approach $0^+$ as $t \to \infty$, i.e. particles are just `kissing' when they stop moving. 
The contact number $z(t\to\infty)$ continuously approaches the isostatic value $z=6$ as $\varphi \to \varphi_J$. The time fluctuations of the data seen at high density for $N=3000$ become very weak in much larger systems, $N=10^6$. In the following, we focus on the results for $N=3000$. 

We also study the relaxation dynamics from steady-shear initial configurations, see Figs.~\ref{fig1}(c,d). In contrast to the random case, the energy decays exponentially $E(t) \sim e^{-t/\tau}$ over the entire time domain, consistent with Ref.~\cite{Olsson2015}.  
The contact number changes very little during the relaxation. This suggests the steady-shear initial configurations (i.e. instantaneous configurations in the steady-state shear flow) have almost the same contact networks as those of the final configurations, and they undergo exponential relaxation from the beginning. 

\begin{figure}[t]
\includegraphics[width=\columnwidth]{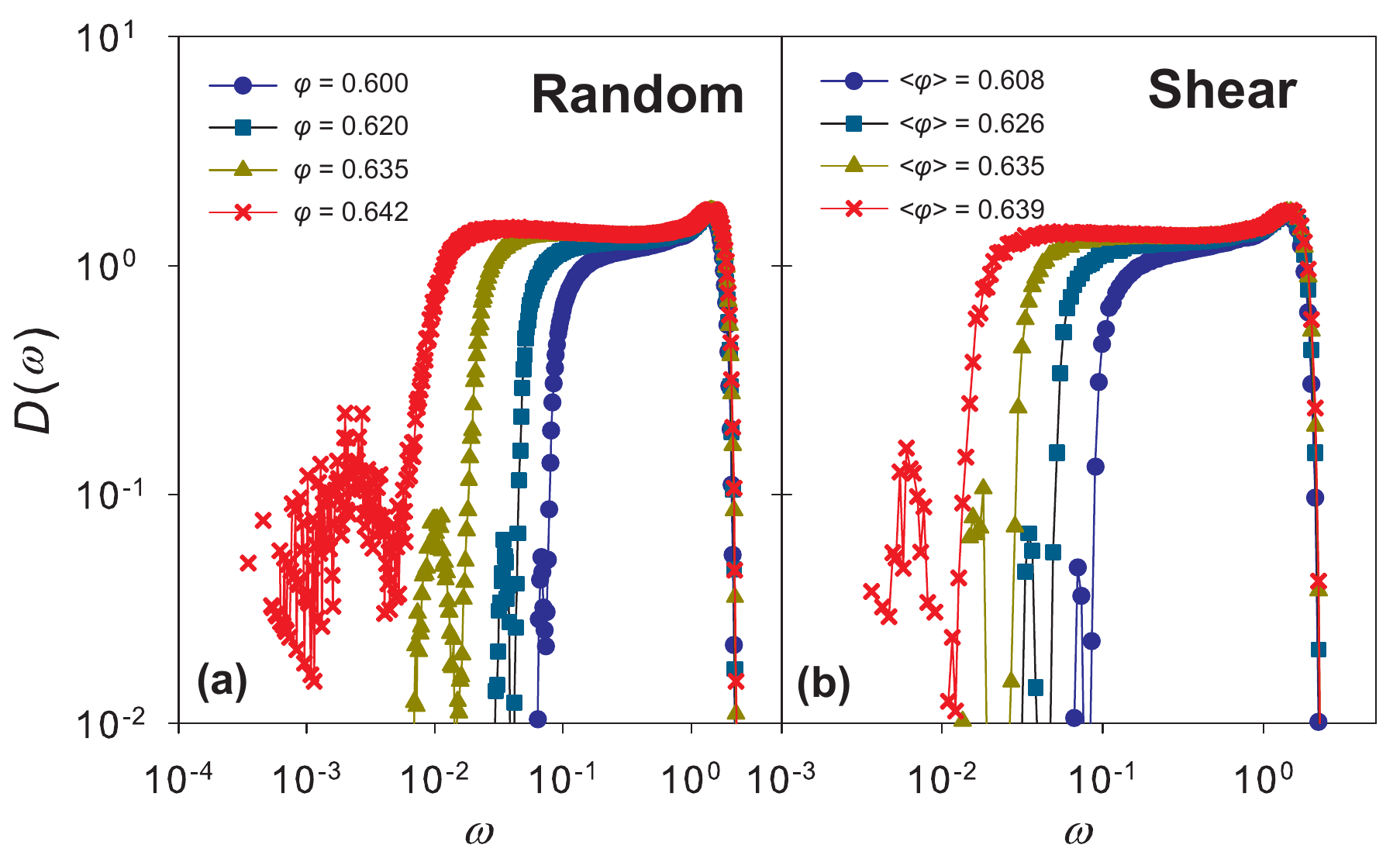}
\caption{
The vibrational density of states $D(\omega)$ for final configurations obtained after relaxation from (a) random and (b) steady-sheared initial configuration. The average over initial configurations with the same density $\varphi$ for (a) and the same shear rate (specified by the average density $\ave{\varphi}$) for (b) broadens the isolated peak at $\omega_{\rm min}$.}
\label{fig2}
\end{figure}

{\em Vibrational density of states.}--- 
We expect that the final exponential decay of the energy towards zero can be controlled by the potential energy landscape around the final configurations. Although the potential energy is nearly zero ($E/N \le 10^{-18}$) in the final configuration, particles still have well-defined contacts, and thus the second derivative of the potential and the Hessian exist. Thus, we can calculate the vibrational density of states (vDOS) of the final configurations. After removing rattlers~\footnote{We remove the particles whose contact numbers are less than 3. We confirmed that, with this choice, the rattler removal does not affect the vDOS. Note that the vDOS changes if we remove the particles whose contact numbers are less than 4.}, we diagonalise the Hessian of each final configuration to obtain the eigenvalues $\lambda_k$ and eigenvectors $\vec{e}_k = [\vec{e}_{1,k}, \vec{e}_{2,k}, \cdots, \vec{e}_{N,k}]$, where $k$ is the mode index. Since the system is hypostatic, we find $N \Delta z/2$ zero modes for each final configuration, as expected from the Maxwell criterion~\cite{Alexander1998}. After removing these zero modes, the vDOSs is calculated as $D(\omega) = \frac{1}{N} \sum_k \delta (\omega - \sqrt{\lambda_k})$. 

In Fig.~\ref{fig2} we show the vDOS for both random and sheared configurations at various densities, averaged over independent final configurations. In both cases, the vDOS has three frequency regimes: a sharp edge at large frequency, a plateau at intermediate frequencies, and an isolated peak at low frequency. With increasing $\varphi$, the plateau and the isolated peak shift to lower frequency and become better separated. The high frequency edge and the intermediate plateau are well-known features of the vDOS seen above jamming, with an onset frequency for the plateau which scales linearly with the excess contact number~\cite{Wyart2005}. Recently, the effective medium theory, which correctly predicts the vDOS of jammed configurations~\cite{Wyart2010}, was extended to the unjammed region $\varphi < \varphi_J$ and predicted $\omega^* \propto \Delta z$~\cite{During2013}. Our data agree with this prediction. A novel feature of the reported vDOS is the isolated low-frequency mode. The corresponding peak appears broad in Fig.~\ref{fig2} because we performed an ensemble average, but we find a unique isolated mode (referred to as ILM) at frequency $\omega_{\rm min}$ in each individual configuration. This ILM was previously observed for sheared non-Brownian hard spheres~\cite{Lerner2012}. 
The ILM we find should be similar to the one detected in sheared hard spheres, and our results show that it can be accessed from the standard analysis of vDOS of soft particles. To our knowledge, the presence of the ILM has not been observed for random configurations before, and our results suggest that it will always be present in unjammed packings in a wide variety of preparation protocols. 

\begin{figure}[t]
\includegraphics[width=\columnwidth]{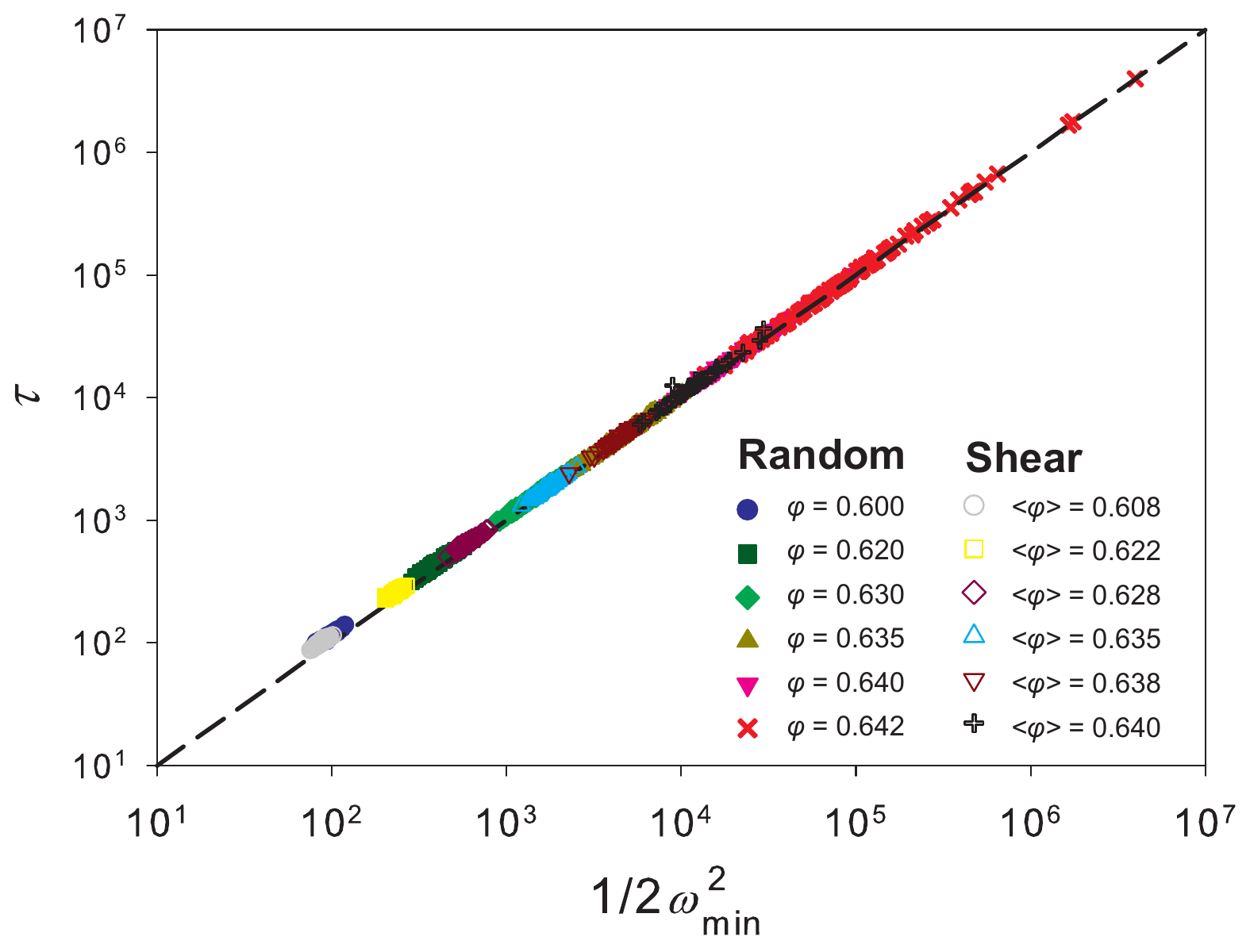}
\caption{Parametric plot of the relaxation time $\tau$ in the exponential regime of the relaxation against $1/(2\omega_{\rm min}^2)$ measured for each individual configuration. The dashed line indicates that $\tau = 1/(2\omega_{\rm min}^2)$ is exactly obeyed for both random (filled symbols) and sheared (open symbols) configurations.}
\label{fig3}
\end{figure}

We now connect the ILM to the relaxation time $\tau$ of the relaxation dynamics. To this end, we parametrically plot $\tau$ against $1/2\omega_{\rm min}^2$ for each individual configuration (the energy relaxation along a mode $k$ follows $e^{-2\lambda_k t}$) in Fig.~\ref{fig3}, which confirms that 
$\tau = 1/2\omega_{\rm min}^2$ is precisely obeyed for configurations at different packing fractions and from different preparation protocol. 
We conclude that the final exponential relaxation of the energy in both random and sheared cases occurs along the ILM. 
We also find that the ILM dominates very strongly over all other modes for the sheared configurations~\footnote{We calculate the overlap between the overall displacements of particles during the relaxation and the eigenvectors of the vibrational modes. We find that the ILM has exceptionally large overlap among all the other modes.}, which presumably explains the absence of a power-law relaxation regime in that case. 

{\em Critical behavior.}--- 
We now focus on the critical behavior of $\omega_{\rm min}$ (and thus of $\tau$) near jamming. To avoid the unwanted effect of sample to sample fluctuations of $\varphi_J$~\cite{Ohern2003}, we plot parametrically in 
Fig.~\ref{fig4} $\omega_{\rm min}$ against $\Delta z$, measured in each individual configuration. Clearly, the two data sets fully overlap without any rescaling over the entire range. Thus, the link between $\omega_{\rm min}$ and $\Delta z$ is universal, independent of the preparation protocol.
This observation implies that the relaxation dynamics from a wide variety of initial configurations below jamming (quantified by $\tau$) is universal, and is directly linked to the shear viscosity $\eta$ of the suspension. We fit the data in Fig.~\ref{fig4} close to the jamming transition ($\Delta z < 0.3$) with a power law $\omega_{\rm min} \approx \Delta z^\alpha$, and find that $\alpha \approx 1.6$ works well. The measured exponent would decrease if we include data for larger $\Delta z$ in the fit. Of course, $\alpha$ could increase a little further if data for even smaller $\Delta z$ could be included. A more precise numerical determination of $\alpha$ is a worthwhile goal for future work, and the vDOS approach used here seems the easiest route for this task.

\begin{figure}[t]
\includegraphics[width=\columnwidth]{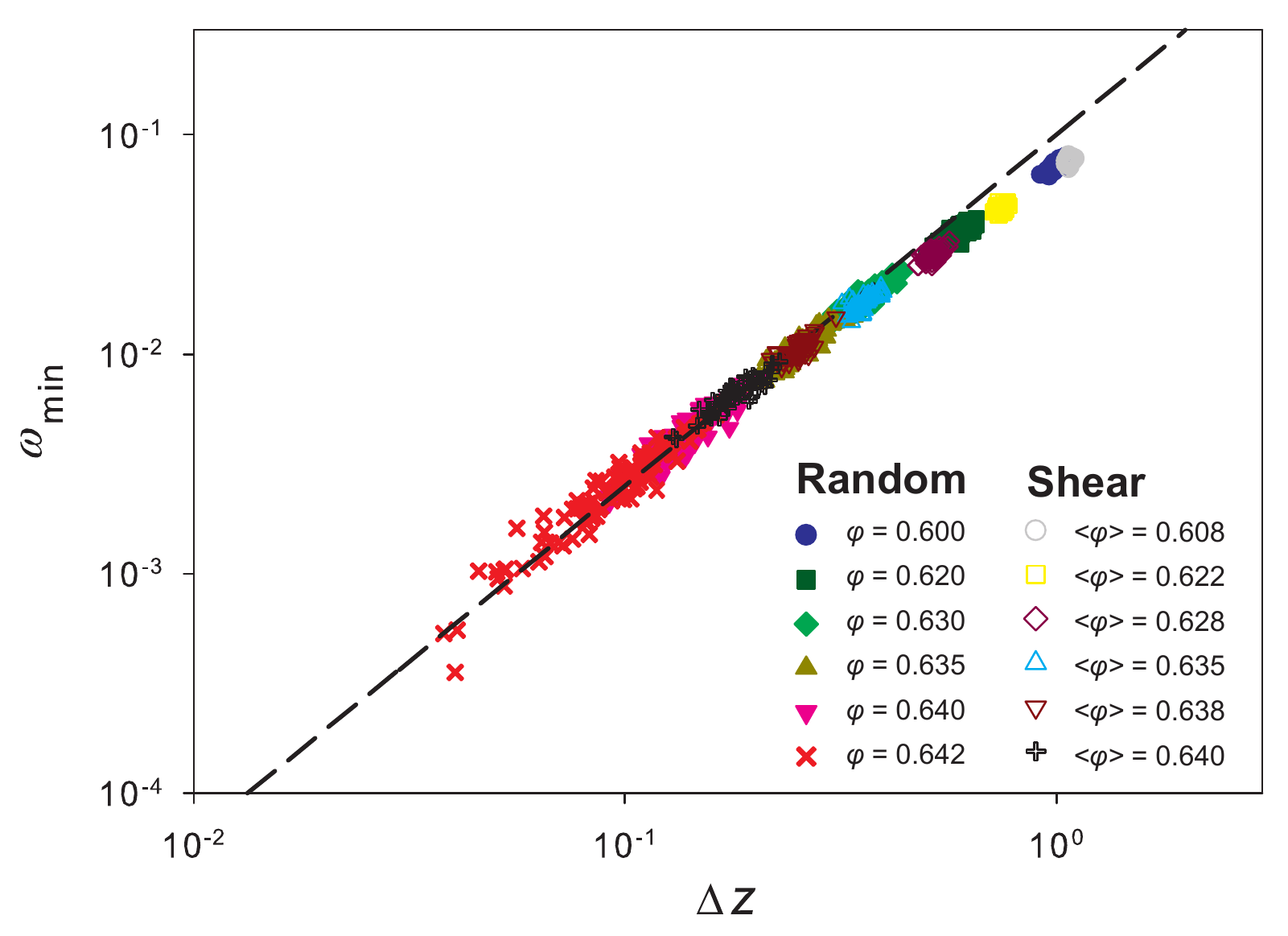}
\caption{Parametric plot of the frequency of the slow mode, $\omega_{\rm min}$, against the distance to isostaticity $\Delta z = 6-z$ for each individual configuration. The dashed line indicates $\omega_{\rm min} \propto \Delta z^{1.6}$, valid at low frequency for both random (filled symbols) and sheared (open symbols) configurations.}
\label{fig4}
\end{figure}

{\em Nature of isolated low-frequency mode.}--- 
Finally, we show that the ILM is qualitatively different from any other mode in the spectrum. Let us consider the evolution of the potential energy obtained by deforming the system along each eigenvector, $\vec{r}_{i} \equiv \vec{r}_{i}^0 + x \vec{e}_{i,k}$, where $\vec{r}_{i}^0$ is the position of particle $i$ in a relaxed configuration, and $x$ is the amplitude of the deformation. We can then track $E=E(x)$ for each mode. We find that $E(x)$ has a parabola shape for all the modes except for the ILM. Fig.~\ref{fig5} shows that $E(x)$ for the ILM is finite for $x<0$, but zero for $x>0$. Namely, the potential energy along the ILM is a one-sided harmonic potential. In addition, the contact number jumps discontinuously from $z(x>0)=0$ to a finite value for $x<0$, which is again different for other modes. This one-sided potential energy landscape connects the mechanical vacuum of unjammed configurations with zero potential energy to the slightly compressed sphere packings obtained when the system is undergoing its relaxational dynamics. The curvature of the potential is given by $\omega_{\rm min}^2$, which thus controls the long-time relaxation dynamics. The direct connection to the shear viscosity stems from the numerical observation that under simple shear, particle displacements are essentially occurring along the ILM. It is remarkable that a single mode captures essentially the entire relaxational physics, rather than a more complex spectrum associated to more complex time dependences as is typically the case in disordered materials~\cite{Berthier2011}. 


\begin{figure}[t]
\includegraphics[width=\columnwidth]{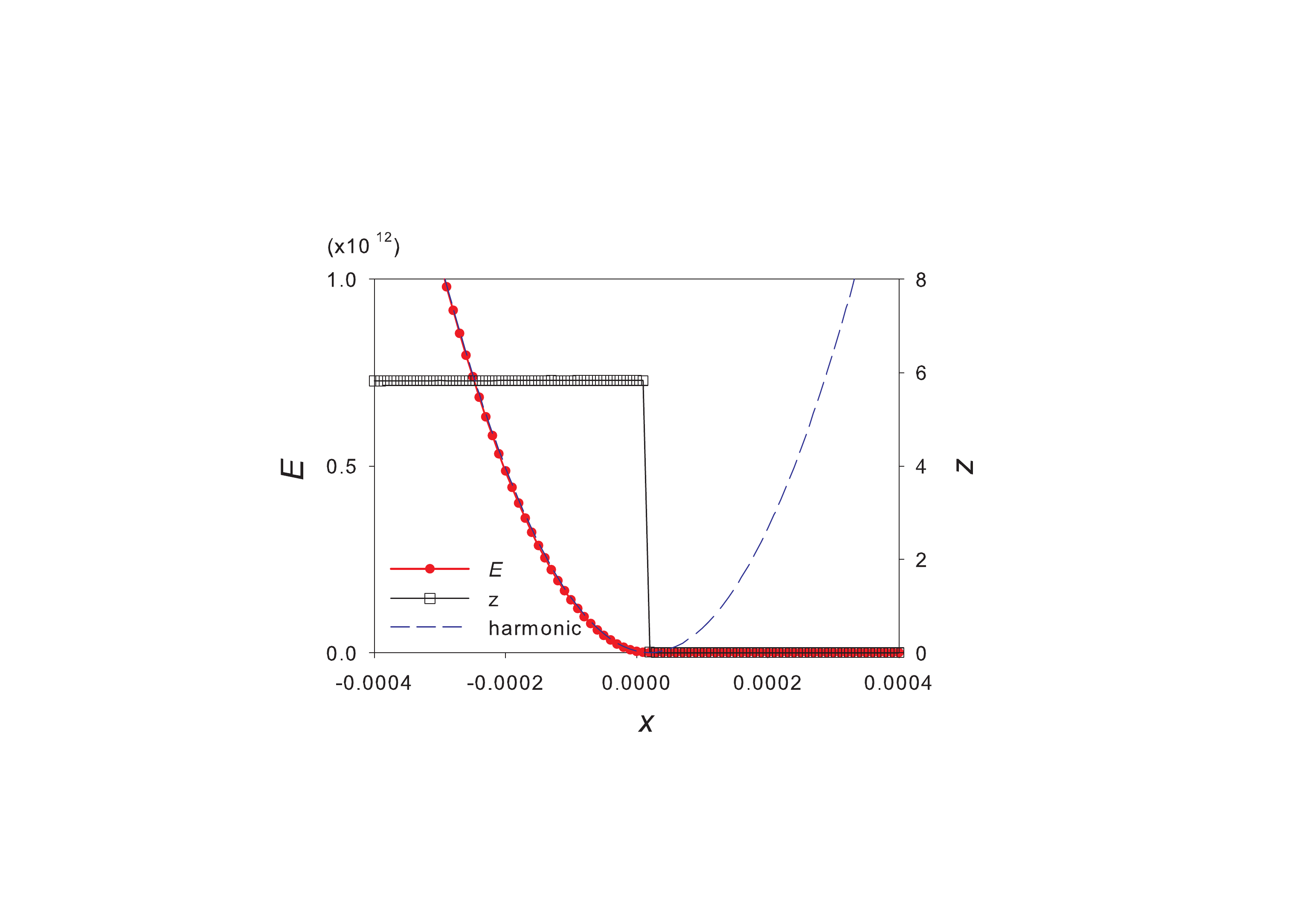}
\caption{The one-sided potential energy landscape along the mode with the lowest frequency $\omega_{\rm min}$ measured for a single configuration at $\varphi = 0.64$. 
It follows the harmonic behavior $E = \frac{1}{2} \omega_{\rm min}^2 x^2$ (indicated by the dashed line) for $x<0$. The discontinuous change of the contact number at $x=0$ is also shown.}
\label{fig5}
\end{figure}

{\em Conclusion.}--- 
We studied the relaxation dynamics of soft particles in the unjammed phase. We considered two types of initial configurations, random and steady sheared.
Despite differences at short time, the relaxation is always exponential at large times, $E \sim \exp(-t/\tau)$. All final configurations display an isolated low-frequency mode at frequency $\omega_{\rm min}$, and the relation $\tau = 1/(2 \omega_{\rm min}^2)$ is accurately obeyed for each individual configuration, independently of their preparation protocol. The final relaxation process thus occurs along the isolated low-frequency mode, which represents the softest mode that can simultaneously and collectively remove all overlaps from the final packing.  Our main quantitative result is that the late stage of the relaxation dynamics is universal, $\omega_{\rm min} \propto \Delta z^{\alpha}$, with $\alpha \approx 1.6$ in $d=3$. 


Our results, on the one hand, establish the universality of the relaxation of athermal particles near the jamming transition. 
Because the relaxation from two extreme types of initial configurations  exhibit the same slowing down, we expect the same in many other physical situations, such as creaming emulsions and injecting foams. 
It would be interesting to study other protocols. 
We are currently studying dynamics after a step shear strain~\cite{Hatano2009} more quantitatively. It was reported that during the relaxation dynamics after a step strain~\cite{Hatano2009}, the stress exhibits a power law decay $\propto t^{-1/2}$ at short times, followed by an exponential decay.
This is similar to our findings with random initial conditions, since stress (and pressure) scale as $E^{1/2}$ for harmonic spheres. On the other hand, our results open a new way to study the viscosity divergence at the jamming transition without applying shear. The exact mean-field solution was recently derived for several properties of jammed configuration, but this approach is still unable to predict the exponents $\alpha$ and $\nu$ discussed in our work. Our results suggest that these critical exponents can be theoretically analysed by studying the simpler problem of the quench dynamics in the unjammed phase without applying a shear flow. 

We thank Hideyuki Mizuno, Masanari Shimada, Matthieu Wyart, and Douglas Durian for useful discussions and comments. 
This work was supported by KAKENHI grants No. 16H04034, 16H04025, 16H06018, 16H06478, 17H04853, 18H01188, 18H05225, and 18K13464, and a Grant from the Simons Foundation (\# 454933, L. B.). Some calculations were performed using the Research Center for Computational Science, Okazaki, Japan.

\end{document}